\documentclass[pra,twocolumn,floatfix,superscriptaddress]{revtex4}
\usepackage{graphicx}
\usepackage{color}
\usepackage[normalem]{ulem}
\usepackage{braket}
\usepackage{amsmath}
\usepackage{verbatim}
\usepackage{amssymb}

\newcommand{\be}{\begin{equation}}
\newcommand{\ee}{\end{equation}}
\newcommand{\ba}{\begin{eqnarray}}
\newcommand{\ea}{\end{eqnarray}}
\newcommand{\nn}{\nonumber \\}

\begin{document}

\title{Ground-state entanglement in coupled qubits}

\author{A.~Yu.~Smirnov }
\affiliation{D-Wave Systems Inc., 3033 Beta Avenue, Burnaby BC
Canada V5G 4M9}
\author{M.~H.~Amin }
\affiliation{D-Wave Systems Inc., 3033 Beta Avenue, Burnaby BC
Canada V5G 4M9} \affiliation{Department of Physics, Simon Fraser
University, Burnaby, BC Canada V5A 1S6}

\begin{abstract}
{We study a system of qubits that are coupled to each other via only
one degree of freedom represented, e.g., by $\sigma_z$-operators. We
prove that, if by changing the Hamiltonian parameters, a
non-degenerate ground state of the system is continuously
transformed in such a way that the expectation values of $\sigma_z$
operators of at least two coupled qubits change, this ground state
is entangled. Using this proof, we discuss connection between energy
level anticrossings and ground state entanglement. Following the
same line of thought, we introduce entanglement witnesses, based on
cross-susceptibilities, that can detect ground state entanglement
for any bipartition of the multi-qubit system. A witness for global
ground state entanglement is also introduced.
 }
\end{abstract}

\maketitle

\section{Introduction}

Entanglement is considered to be an essential property required for
quantum computation \cite{Guhne09,Amico08}. It has been shown that a
pure state quantum computation that does not possess some minimum
level of entanglement can be efficiently simulated by classical
computers \cite{Vidal03,Jozsa03}.


For a pure state, entanglement can be defined in the following way
\cite{Guhne09}. Consider an arbitrary
bipartition of the system $S$ into two parts $A$ and $B$, with
Hilbert spaces ${\cal H}_S$ and ${\cal H}_{A,B}$, respectively. The
system $S$ described by a wave function $\ket{\Psi}\in {\cal H}_S$
is entangled if there exists at least one bipartition $A$ and $B$
such that $\ket{\Psi}$ cannot be represented as a tensor product
$\ket{\Psi_A} {\otimes} \ket{\Psi_B}$ of the states $\ket{\Psi_A}\in
{\cal H}_A$ and $\ket{\Psi_B}\in {\cal H}_B$. By this definition, if
an $N$ qubit system is not entangled, then it is completely
separable: $\ket{\Psi} {=} {\otimes}_{i=1}^N \ket{\psi_i}$, where
$\ket{\psi_i}$ is a single qubit wave function in the Hilbert space
of qubit $i$. One can also define entanglement for a particular
bipartition when $\ket{\Psi}$ is not separable relative to that
partition.  Global entanglement is defined when $\ket{\Psi}$ is not
separable for any bipartition.


Entanglement is usually characterized by measures and witnesses
\cite{Guhne09,Wootters98,VidalWerner02,MW02,Brennen03,Love07,Spedalieri12}.
A measure of entanglement provides a quantitative characterization
for the amount of entanglement. A witness of
entanglement, on the other hand, provides a sufficient but not necessary condition for entanglement. Usually, the condition is written in a form of an inequality. If the inequality is satisfied, the system is entangled,
but if it does not, the system may or may
not be entangled. Witnesses are constructed in such a way that they
can be measured in practice \cite{Neeley10,DiCarlo10}.

Experimentally, entanglement is commonly detected by quantum state
tomography  \cite{Steffen06} or by measuring correlations between
different parts of the system that are not allowed by classical
physics \cite{GHZ,Ansmann09}. Most experiments involve preparing the
system in a particular state by turning on interactions between the
subsystems, but performing measurements when there is no interaction
between the qubits. This is different from detecting entanglement
when the interaction between the subsystems is always present.
Detecting entanglement in that case, for example, when the system is
in an eigenstate of a Hamiltonian, requires a different approach.
Any such approach is inherently restricted by experimental
limitations: the ability to measure qubits in a particular basis,
limitations in time domain control, etc.

Spectroscopy is one of the tools that has been used to detect
coherence and entanglement in multi-qubit systems
\cite{Friedman00,Berkley03,Berkley13}. In these measurements,
observation of an anticrossing between two energy levels is taken to
be an evidence for eigenstate superposition or entanglement. While
this is intuitively clear, to our knowledge there is no mathematical
justification to support it. In this paper, we prove a theorem
that relates anticrossings with ground state entanglement
for a system of qubits with pairwise coupling via a single degree of
freedom.

Magnetic susceptibility has also been used to detect entanglement
\cite{Ghosh03}. An entanglement witness using a sum of
susceptibilities taken along $x$-, $y$-, and $z$-directions has been
proposed in Ref.~\cite{Wiesniak05} and measured in
Refs.~\cite{Souza08,Soares09}. The witness requires measurements of
average values of spin projections in all three dimensions. This
becomes problematic when only one component of the spin projection
can be measured, unless the system is isotropic \cite{Brukner06}.
Here, we introduce witnesses, based on only one component of the
 susceptibility, to detect ground state entanglement. No
detailed knowledge of the Hamiltonian, except that the qubits are
pairwise coupled via $\sigma_z$-operators, is needed. These
witnesses can provide useful tools for experimental demonstration of
entanglement in a quantum annealing processor
\cite{Harris10,MarkNature11}.

The paper is organized as follows.  In Sec.~II we prove that
anticrossing of energy levels can be considered as a signature of
entanglement. Susceptibility-based entanglement witnesses are
described in Sec.~III and conclusions are provided in Sec.~IV.

%


\section{Anticrossings and entanglement}

Observation of avoided crossings or anticrossings between two energy
levels have been used as evidence for coherent superposition
\cite{Friedman00} and entanglement \cite{Berkley03, Karthik07}. It
is intuitively evident that at the center of an avoided crossing the
eigenstates of the many-qubit system are superpositions of the
two crossing states and therefore can be entangled. If all terms in
the Hamiltonian are known, one can easily check by exact
diagonalization whether or not the eigenstates are entangled. But
without the exact knowledge of the Hamiltonian, is it possible to
conclude anything about the entanglement?

Consider a system of qubits described by the following Hamiltonian
\begin{eqnarray}
H= \sum_{i=1}^N H_i + {1\over 2}\sum_{i,j=1}^N J_{ij}
\sigma^z_i\sigma^z_j, \label{HS}
\end{eqnarray}
where $N$ is the number of qubits, $H_i$ is a single qubit
Hamiltonian acting on the $i$-th qubit, $J_{ji}=J_{ij}$, and
$J_{ii}=0$. Operators $\sigma^z_i$ acting in the many-qubit Hilbert
space are defined as
 \be
 \sigma^z_i \equiv \otimes_{k=1}^N \hat o_k,  \qquad
 \hat o_k = \left\{ \begin{array}{cc}
 \sigma_z &  \text{ if } k=i \\
 I & \text{ if } k\ne i \end{array} \right.
 \label{sigmaz}
 \ee
where $I$ is a $2{\times}2$ identity matrix and $\sigma_z$ is a
standard Pauli matrix, $\sigma_z = \bigl(\begin{smallmatrix} 1&0\\
0&-1 \end{smallmatrix} \bigr)$. We assume no knowledge of $H_i$ and
the exact values of $J_{ij}$. The only knowledge is that the qubits
are coupled via $\sigma^z_i$-operators. In general, $H$ is a
function of a set of external parameters which we collectively
denote by $\lambda$. These parameters are, for example, external
voltages, currents, or fluxes applied to the physical qubits.

Let $\ket{\Psi_0(\lambda)}$ be the ground state of $H(\lambda)$,
and $\braket{\sigma^z_j} \equiv \braket{\Psi_0|\sigma^z_j|\Psi_0}$. \\

\noindent {\bf Theorem:} \emph{Consider Hamiltonian (\ref{HS}) that
has at least one nonzero coupling and is a continuous function of
$\lambda \in \Lambda$, where $\Lambda$ is a connected region in the
parameter space. If $\ket{\Psi_0(\lambda)}$ is a completely
separable non-degenerate ground state of $H(\lambda)$, then for
every pair $\{i,j\}$ of qubits with $J_{ij}\neq 0$, there is at
least one average spin projection, $\braket{\sigma^z_i}$ or
$\braket{\sigma^z_j}$ or both,
which does not change with $\lambda$.} \\

\noindent We provide a proof for this theorem in appendix A. Here,
we only focus on its implications. The theorem states that if there
is at least one pair of coupled qubits $i$ and $j$ (with $J_{ij}
\neq 0$) for which the expectation values $\braket{\sigma^z_i}$ and
$\braket{\sigma^z_j}$ of \emph{both} qubits simultaneously change
when the Hamiltonian parameters change, and if the system stays in a
unique eigenstate during this change, then that eigenstate should be
entangled.


The best example is when the ground state of the system goes through
an anticrossing as the parameters are varied. Consider the
transverse Ising Hamiltonian
\begin{equation} \label{Ham1}
H_{\rm TI} = - \frac{1}{2} \sum_i \Delta_i \sigma^x_i - \sum_i h_i
\sigma^z_i + \sum_{i < j} J_{ij} \sigma^z_i \sigma^z_j,
\end{equation}
where $\sigma^x_i$ are defined similar to $\sigma^z_i$ in
(\ref{sigmaz}).  We focus on a system of ferromagnetically coupled
qubits with $J_{ij}=-J$, all subject to a uniform energy bias $h_i =
h$. When $h = - h_0 < 0 $ is a large negative number, the state of
the system will be close to the ferromagnetically ordered state
$\ket{\downarrow \downarrow ... \downarrow}$. On the other hand for
large positive values, $h = h_0
> 0$, the state of the system will be close to $\ket{\uparrow
\uparrow ... \uparrow}$. In this example, $h$ plays the role of
$\lambda$. Clearly at $\lambda_{\rm in} = h_{\rm in}=-h_0$, all
qubits have $\braket{\sigma^z_i(\lambda_{\rm in})} \approx - 1$ in
the ground state and at $\lambda_{\rm fin} = h_{\rm fin}=h_0$, all
qubits have $\braket{\sigma^z_i(\lambda_{\rm fin})} \approx 1$.
Therefore the ground state expectation value of $\sigma^z_i$ changes
for all qubits. Now, if in changing from $\lambda_{\rm in}$ to
$\lambda_{\rm fin}$ the system goes through an anticrossing, it
means that one can continuously change the ground state of the
system without going through any degeneracy, i.e., the ground state
remains non-degenerate all through the change. In that case, the
above theorem states that the ground state of the system has to go
through an entangled state during this evolution unless all qubits
are uncoupled ($J_{ij}=0, \ \forall\, i,j$). The latter can be
easily checked if instead of applying a uniform bias to all qubits,
one applies a bias to only one of the qubits. In that case, if the
qubits are uncoupled, only $\braket{\sigma^z_i}$ of the qubit to
which the bias is applied will change and the other qubits remain
unaffected. Therefore, if by applying a bias to one qubit, the
expectation value $\braket{\sigma^z_i}$ for all other qubits get
affected, then the qubits should be coupled. Observation of such an
anticrossing, therefore, is evidence for the existence of
entangled ground state. Notice that knowledge of the exact values of
$J_{ij}$ or any other Hamiltonian parameters is not necessary to
prove ground state entanglement.

\section{Susceptibility-based entanglement witnesses}

The theorem in the previous section suggests a close relation
between entanglement and  susceptibility. Let us define
susceptibility of qubit $i$ to $\lambda$ as
 \be
 \chi_{i}^\lambda = \partial \braket{\sigma^z_i}/\partial \lambda.
 \ee
The theorem states that for two coupled
qubits, $J_{ij}\ne 0$, if both susceptibilities $\chi_i^\lambda$ and
$\chi_j^\lambda$ are nonzero, then the state is entangled. This
suggests introducing
 \be
 {\cal W}_\lambda = \sum_{ij} |J_{ij}\chi_i^\lambda\chi_j^\lambda|,
 \ee
which is zero if the state is completely separable. A nonzero value
of ${\cal W}_\lambda$ means that at least for one pair of qubits, all
three $J_{ij}$, $\chi_i^\lambda$, and $\chi_j^\lambda$ are nonzero
and therefore the system is entangled. The inequality ${\cal W}_\lambda > 0$, therefore, is a sufficient condition for entanglement. As such, we consider ${\cal W}_\lambda$ as an entanglement witness, based on our convention. The above witness, however,
cannot determine how many qubits are entangled to each other. Even
if only two qubits are entangled, the witness will be nonzero.

One can generalize the above idea to introduce witnesses that can
distinguish any bipartite entanglement. For the rest of this section
we focus on the transverse Ising Hamiltonian (\ref{Ham1}), for which
we can define the ground state cross-susceptibility between qubit
$i$ and qubit $j$ as
\begin{equation} \label{ChiDef}
\chi_{ij} = \frac{ \partial \braket{\sigma^z_i}}{\partial h_j}\;,
\end{equation}
For a Hamiltonian $H$ with eigenstates $\ket{\Psi_{n}}$  and
eigenvalues $E_{n}$, the susceptibility can be written as (see
Ref.~\cite{VanVleck78} and also Appendix B)
\begin{eqnarray} \label{Chi1}
\chi_{ij} &=& \sum_{n > 0} \frac{ \braket{\Psi_0 |\sigma^z_{j}
|\Psi_{n} } \braket{\Psi_{n} |\sigma^z_{i} |\Psi_{0} }} {E_n - E_0}
+ c.c.
\end{eqnarray}
Here, $c.c.$ means the complex conjugate of the previous expression.

Consider an arbitrary bipartition of the system $S$ into two parts
$A$ and $B$, $S {=}\, A {\cup} B$. Recall that these two parts are
separable if the wave function of $S$ can be written as
$\ket{\Psi_0} = \ket{\Psi_0^A}{\otimes}\ket{\Psi_0^B}$, otherwise,
they are entangled. Our first goal is to define a witness that can
detect entanglement between $A$ and $B$ when $\ket{\Psi_0}$ is a
non-degenerate ground state of $H$. Let us introduce
 \ba
 \widetilde{\cal W}_{AB} = \sum_{i\in A,j\in B} J_{ij} \chi_{ij}.
 \ea
Using (\ref{Chi1}), we can write
 \ba
 && \widetilde{\cal W}_{AB} =\sum_{n > 0} {1 \over E_n - E_0} \nn
 && \ \times \sum_{i\in A,j\in B} J_{ij} \braket{\Psi_0 |\sigma^z_{j} |\Psi_{n} }
\braket{\Psi_{n} |\sigma^z_{i} |\Psi_{0} } + c.c. \label{WAB}
 \ea
It is clear that $\widetilde{\cal W}_{AB}$ is not only a function of
the ground state, $\ket{\Psi_0}$, but also a function of all other
eigenstates. However, as we shall see, it can still be used to
detect the ground state entanglement.
%

Let us assume that the ground state is separable, $\ket{\Psi_0} =
\ket{\Psi_0^A}{\otimes}\ket{\Psi_0^B}$. We introduce two reduced
vectors,
\begin{equation}
\ket{\widetilde{\Psi}_{n}^A } = \braket{\Psi_0^B | \Psi_{n}} \in
{\cal H}_A, \quad \ket{\widetilde{\Psi}_{n}^B } = \braket{\Psi_0^A |
\Psi_{n}} \in {\cal H}_B.
\end{equation}
Due to orthogonality of the ground and excited states,
$$
\braket{\widetilde{\Psi}_{n}^A  | \Psi_0^A} = 0, \qquad
\braket{\widetilde{\Psi}_{n}^B  | \Psi_0^B} = 0. $$ By definition,
$\ket{\Psi_0}$ is an eigenstate of (\ref{Ham1}), thus
\begin{equation} \label{EqH}
(H_A + H_B + \sum_{i\in A,j \in B} J_{ij} \sigma^z_i \sigma^z_j )
\ket{\Psi_0}  = E_0 \ket{\Psi_0}.
\end{equation}
Here, $H_A$ and $H_B$ describe the subsystems $A$ and $B$,
respectively. We can multiply Eq.~(\ref{EqH}) by
$\bra{\widetilde{\Psi}_{n}^B}$ to obtain
\begin{eqnarray} \label{EqS1}
\sum_{i\in A,j \in B} J_{ij} \braket{\widetilde{\Psi}_{n}^B |
\sigma^z_j | \Psi_0^B} \sigma^z_i \ket{\Psi_0^A} = -
\braket{\widetilde{\Psi}_{n}^B | H_B | \Psi_0^B} \ket{\Psi_0^A}.
\nonumber
\end{eqnarray}
Multiplying by  ${\otimes}\ket{\Psi_0^B}$ and using the definition
of $\ket{\widetilde{\Psi}_{n}^B}$, we find
\begin{eqnarray} \label{EqS2}
\sum_{i\in A,j \in B} J_{ij} \braket{{\Psi}_{n} | \sigma^z_j |
\Psi_0} \sigma^z_i \ket{\Psi_0} = - \braket{\widetilde{\Psi}_{n}^B |
H_B | \Psi_0^B} \ket{\Psi_0}. \nonumber
\end{eqnarray}
Taking complex conjugate of all terms we get
\begin{eqnarray} \label{EqS2}
\sum_{i\in A,j \in B} J_{ij} \braket{\Psi_0 | \sigma^z_j | \Psi_n}
\sigma^z_i \ket{\Psi_0}^* = - \braket{\Psi_0^B| H_B
|\widetilde{\Psi}_{n}^B } \ket{\Psi_0}^*, \nonumber
\end{eqnarray}
where $\ket{\Psi_0}^*$ is a vector with each component being the
complex conjugate of the corresponding component in  $\ket{\Psi_0}$.
If the Hamiltonian is real in the computation basis, $H = H^*,$
i.e., only contains $\sigma^x_i$ and $\sigma^z_i$ matrices with real
coefficients, then $H \ket{\Psi_0}^* = E_0 \ket{\Psi_0}^*$.
Therefore, $\ket{\Psi_0}$ and $\ket{\Psi_0}^*$ should be
proportional to each other, $\ket{\Psi_0}^* = e^{i \varphi_0}
\ket{\Psi_0},$ (up to a constant phase factor $e^{i\varphi_0})$
provided that the ground energy level $E_0$ is non-degenerate.
Replacing  $\ket{\Psi_0}^*$ with $\ket{\Psi_0}$ and applying
$\bra{\Psi_n}$ from the left, we get \be \sum_{i\in A,j \in B}
J_{ij} \braket{{\Psi}_0 | \sigma^z_j | \Psi_n}
\braket{\Psi_n|\sigma^z_i|\Psi_0} = 0. \ee Comparing with the second
line of (\ref{WAB}), we find $\widetilde{\cal W}_{AB} = 0$.
Therefore, if the measurements show that $\widetilde{\cal W}_{AB}
\ne 0$, this means that $\ket{\Psi_0}$ is not separable relative to
$A$ and $B$ and therefore is entangled. This is true independent of
whether the excited states are separable or entangled.

Since susceptibility can be divergent, $\widetilde{\cal W}_{AB}$ is
not bounded. To have a witness that is bounded by 1 and is
independent of the number of the connections between the subsystems
$A$ and $B$, it is useful to introduce
\begin{equation}
{\cal W}_{AB} = \frac{|\widetilde{\cal W}_{AB}
|}{N_{AB}+|\widetilde{\cal W}_{AB} |} , \label{WAB1}
\end{equation}
where $N_{AB}$ is the number of non-zero couplings between
subsystems $A$ and $B$. From the above argument, a nonzero value of
${\cal W}_{AB}$ means that the two parts $A$ and $B$ are entangled
in the ground state of the transverse Ising Hamiltonian.

The witness ${\cal W}_{AB}$ can only determine entanglement between
$A$ and $B$. It cannot determine whether every part of $S$ is
entangled with every other part. To detect the global entanglement,
one can use geometric averages \cite{Love07}. Suppose that there are
$N_p$ possible bipartitions of $S$. A global entanglement witness
can be defined as
\begin{equation}
{\cal W}_{\chi} = \frac{\left[\prod (\widetilde{\cal
W}_{AB}/N_{AB})\right]^{1/N_p}}{ 1 + \left[\prod (\widetilde{\cal
W}_{AB}/N_{AB})\right]^{1/N_p}}.
\end{equation}
where $\prod$ denotes the product over all $N_P$ possible
bipartitions of $S$. Similar to (\ref{WAB1}), ${\cal W}_{\chi}$ is
bounded by 1. A nonzero value of ${\cal W}_{\chi}$ means that all
parts of $S$ are entangled to each other, therefore, the system is
globally entangled.

\section{Conclusions}

Ground state entanglement in a system of qubits that are pairwise
coupled to each other via a single degree of freedom has been
studied. We have presented a theorem that relates eigenstate
entanglement with anticrossing of the energy levels. We have
introduced ground state entanglement witnesses, based on
susceptibility, which can detect any bipartite entanglement as well
as the global entanglement. This approach enables experimental
demonstration of entanglement of coupled qubits without detailed
information of the system Hamiltonian.

In order to apply these results in experiments, one needs to have
access to spectroscopy, a method to measure ground state average of the spin
projection operator, $\braket{\sigma^z_i}$, and a way to measure coupling
strengths or at least determine if they are nonzero. Information
about the dependence of $\braket{\sigma^z_i}$ on the Hamiltonian
parameters is important for the proof of entanglement via
anticrossings as well as for the extraction of the
susceptibilities used in the entanglement witnesses. Usually, in experiments $\sigma^z_i$ is measured in the thermal state. To assure that
the measurement provides a good
approximation for the ground state average  $\braket{\sigma^z_i}$,
it is important to have a pronounced energy gap, in the
spectroscopic data, between the ground and first excited states. The
gap has to be larger than the energy level broadening and
temperature to guarantee that the ground state remains the only
thermally occupied state during the measurement.

\section*{Acknowledgements}
We are grateful to S. Boixo, F. Spedalieri, T. Lanting, and P. Love
for helpful discussions.

\appendix

\section{Proof of the theorem}

In this appendix, we provide a proof for the theorem presented in Sec.~II. First, we prove a few Lemmas.  Consider Hamiltonian $H$ defined by (\ref{HS}). \\

\noindent \textbf{Lemma 1:} \emph{If $H$ has a completely separable
eigenstate $\ket{\Psi} = \otimes_{i=1}^N \ket{\psi_i}$, where
$\ket{\psi_i}$ is a state in the Hilbert space of the $i$-th qubit,
then for each nonzero coupler, $J_{ij} \neq 0$, either
$\ket{\psi_i}$ or $\ket{\psi_j}$ is an eigenfunction of the
$\sigma_z$-operator: $\sigma_z \ket{\psi_i} = z_i \ket{\psi_i}$ or
$\sigma_z \ket{\psi_j} = z_j \ket{\psi_j}$, with corresponding
eigenvalues $z_i$ or $z_j$ ($z_i\in \{0,1\}$).} \\

If $\ket{\Psi}$ is an eigenstate of $H$, then
\begin{equation} \label{L11}
H\ket{\Psi} = E \ket{\Psi}
\end{equation}
We multiply both sides of (\ref{L11}) by $({\otimes}_{k\ne i,j}
\bra{\psi_k}){\otimes} \bra{\widetilde\psi_j}$, where
$\ket{\widetilde\psi_j}$ is a state orthogonal to $\ket{\psi_j}$:
$\bra{\widetilde\psi_j}\psi_j\rangle = 0$. Using (\ref{HS}) we find
 \begin{equation} \label{L12}
 J_{ij}\bra{\widetilde\psi_j}\sigma_z\ket{\psi_j} \sigma_z\ket{\psi_i} = E_i(\widetilde\psi_j)\ket{\psi_i}
 \end{equation}
where
 \begin{eqnarray}
 E_i(\widetilde\psi_j) &=& {-} \bra{\widetilde\psi_j}H_j\ket{\psi_j} \nonumber\\  && {-} \sum_{k\ne i,j} J_{kj}
 \bra{\psi_k}\sigma_z\ket{\psi_k}\bra{\widetilde\psi_j}\sigma_z\ket{\psi_j} \label{L13}
 \end{eqnarray}
Since $J_{ij} \ne 0$, Eq.~(\ref{L12}) requires $\ket{\psi_i}$ to be
an eigenfunction of $\sigma_z$ unless if
$\bra{\widetilde\psi_j}\sigma_z\ket{\psi_j}=0$. The latter, however,
requires $\ket{\psi_j}$ to be an eigenfunction of $\sigma_z$. To see
this, we notice that the orthogonal set $\{
\ket{\psi_j},\ket{\widetilde\psi_j}\}$ forms a complete set of bases
in the Hilbert space of qubit $j$. One can therefore write
$\sigma_z\ket{\psi_j} = c_1 \ket{\psi_j} + c_2
\ket{\widetilde\psi_j}$. If
$\bra{\widetilde\psi_j}\sigma_z\ket{\psi_j}=0$, then $c_2=0$ and
therefore $\ket{\psi_j}$ should be an eigenfunction of $\sigma_z$
with eigenvalue $c_1$ ($=\pm1$). This means that at least one of
$\ket{\psi_i}$ and $\ket{\psi_j}$ is an eigenfunction of $\sigma_z$.
$\square$

From (\ref{L12}) and (\ref{L13}), it also follows that if
$\ket{\psi_j}$ is an eigenfunction of $\sigma_z$, then
$E_i(\widetilde\psi_j) = {-}
\bra{\widetilde\psi_j}H_j\ket{\psi_j}=0$. Therefore, $H_j$ should
also be diagonal in the basis $\{
\ket{\psi_j},\ket{\widetilde\psi_j}\}$.\\

\noindent \textbf{Lemma 2:} {\em Consider a state $\ket{\Psi} =
\ket{z_i} {\otimes} \ket{\Phi^i}$, where $\ket{z_i}$  is an
eigenfunction of $\sigma_z$  ($z_i\in \{0,1\}$) representing the
state of qubit $i$, and $\ket{\Phi^i}$ is a state in the Hilbert
space, $S_i$, of all qubits except the $i$th one. If $\ket{\Psi}$ is
an eigenstate of Hamiltonian (\ref{HS}), then all eigenstates of $H$
can be represented as either $\ket{\Psi_n} = \ket{z_i} {\otimes}
\ket{\Phi_n^i}$ or $\ket{\bar \Psi_n} =\ket{\bar z_i} {\otimes}
\ket{\bar \Phi_n^i}$, where $\bar z_i = 1-z_i$, and
$\{\ket{\Phi_n^i}\}$ and $ \{\ket{\bar \Phi_n^i}\}$, with
$n{=}1,...,2^{N-1}$, are two independent complete orthogonal sets in
$S_i$.} \\

Multiplying both sides of (\ref{L11}) by $\bra{\Phi} {\otimes}
\bra{\bar z_i}$ and using $\bra{\bar z_i}\sigma_z\ket{z_i}=0$, we
find $\bra{\bar z_i}H_i\ket{z_i} = 0$. This means that $H_i$ is
diagonal in the basis $\{ \ket{0},\ket{1}\}$, i.e.,
$H_i\ket{z_i}=E_i\ket{z_i}$  and $H_i\ket{\bar z_i}=\bar
E_i\ket{\bar z_i}$. Defining $\ket{\Psi_n} = \ket{z_i} {\otimes}
\ket{\Phi_n^i}$ and $\ket{\bar \Psi_n} =\ket{\bar z_i} {\otimes}
\ket{\bar \Phi_n^i}$, with $n{=}1,...,2^{N-1}$, and using
$\sigma_z\ket{z_i}=\alpha_i\ket{z_i}$, where $\alpha_i \equiv
2z_i-1$ ($= \pm1$), we find
 \begin{eqnarray}
  H\ket{\Psi_n} = \ket{z_i} {\otimes} H_\Phi^i\ket{\Phi_n^i},
  \nonumber\\
H\ket{\bar \Psi_n} = \ket{\bar z_i} {\otimes} \bar
H_\Phi^i\ket{\bar\Phi_n^i}, \label{A4}
\end{eqnarray}
where
\begin{eqnarray} \label{HPhi1}
H_\Phi^i = E_i + \sum_{j\neq i} ( H_j + \alpha_i J_{ij} \sigma^z_j)
+ \frac{1}{2} \sum_{j,k \neq i} J_{jk} \sigma^z_k \sigma^z_j, \\
\label{HPhi2} \bar H_\Phi^i = \bar E_i + \sum_{j\neq i} ( H_j -
\alpha_i J_{ij} \sigma^z_j) + \frac{1}{2} \sum_{j,k \neq i} J_{jk}
\sigma^z_k \sigma^z_j.
\end{eqnarray}
From (\ref{A4}), it is clear that if $\ket{\Phi_n^i}$ is an
eigenfunction of $H_\Phi^i$ with eigenvalue $E_n$ and $\ket{\bar
\Phi_n^i}$ is an eigenfunction of $\bar H_\Phi^i$ with eigenvalue
$\bar E_n$, then $H\ket{\Psi_n} = E_n\ket{\Psi_n}$ and $H\ket{\bar
\Psi_n} = \bar E_n\ket{\bar \Psi_n}$. Since $\{\ket{\Psi_n}\}$ and
$\{\ket{\bar \Psi_n}\}$ are orthogonal to each other, and there are
$2^{N-1}$
states in each set, all $2^N$ eigenstates of $H$ are in these two sets. $\square$ \\

\noindent {\bf Proof of the theorem:} Assume that $H(\lambda)$ has a
completely separable ground state $\ket{\Psi_0(\lambda)} =
\otimes_{k=1}^N \ket{\psi_k(\lambda)}$. We consider a pair $\{i,j\}$
of connected qubits with $J_{ij} \neq 0.$ It follows from Lemma 1
that at least one of the two states, $\ket{\psi_i}$ or
$\ket{\psi_j}$, should be an eigenstate of the $\sigma_z$-matrix. We
consider two points, $\lambda_L = \lambda_0 - \delta \lambda/2$ and
$\lambda_R = \lambda_0 + \delta \lambda/2$, located in the close
proximity of an arbitrary point $\lambda_0 \in \Lambda$. Without
loss of generality, we take $\ket{\psi_i}$ be a
$\sigma_z$-eigenstate at $\lambda_L$: $\ket{\psi_i (\lambda_L)}  =
\ket{z_i}$. Based on Lemma 1, either $\ket{\psi_i (\lambda_R)}$ or $\ket{\psi_j (\lambda_R)}$ should be a $\sigma_z$-eigenstate. There are three possibilities:\\

\noindent (i) $\ket{\psi_i (\lambda_R)}= \ket{z_i}$. Thus, the
direction of the $i$-th spin does not change during the
infinitesimal transition from $\lambda_L$ to $\lambda_R:$
$\braket{\sigma^z_i}_{\lambda =\lambda_L}  = \braket{\sigma^z_i}_{\lambda =\lambda_R}  = z_i.$\\

\noindent (ii) $\ket{\psi_i (\lambda_R)} = \ket{\bar z_i}$ with
$\bar z_i = 1 {-} z_i$. The flipping of the $i$th qubit between
$\lambda_L$ and $\lambda_R$ requires a crossing of energy levels, in
contradiction to the theorem conditions. To see this, let us write
\begin{eqnarray} \label{PsiLR0}
\ket{\Psi_0 (\lambda_L)} &=& \ket{z_i}\otimes
\ket{\Phi^i(\lambda_L)}, \nonumber\\ \ket{\bar \Psi_0 (\lambda_R)}
&=& \ket{\bar z_i}\otimes \ket{\bar \Phi^i(\lambda_R)},
\end{eqnarray}
where $\ket{\Phi^i(\lambda_L)}$ and $\ket{\bar \Phi^i(\lambda_R)}$
are eigenstates of $H_{\Phi}^i(\lambda_L)$ and $\bar
H_{\Phi}^i(\lambda_R)$ defined in (\ref{HPhi1}) and (\ref{HPhi2}):
\begin{eqnarray} \label{HLR1}
H_{\Phi}^i(\lambda_L)\ket{\Phi^i(\lambda_L)} = E_0(\lambda_L)
\ket{\Phi^i(\lambda_L)}, \nonumber\\
 \bar H_{\Phi}^i(\lambda_R)\ket{\bar
\Phi^i(\lambda_R)} = \bar E_0(\lambda_R) \ket{\bar
\Phi^i(\lambda_R)}.
\end{eqnarray}
Since (\ref{PsiLR0}) are the ground states of (\ref{HS}) at the two
points, we have
\begin{eqnarray}
H \ket{\Psi_0 (\lambda_L)} = E_0(\lambda_L) \ket{\Psi_0
(\lambda_L)}, \nonumber\\
H \ket{\bar \Psi_0 (\lambda_R)} = \bar E_0(\lambda_R) \ket{\Psi_0
(\lambda_R)}.
\end{eqnarray}
According to Lemma 2 the following states,
\begin{eqnarray} \label{PsiLR1}
\ket{\bar \Psi_e (\lambda_L)} &=& \ket{\bar z_i}\otimes \ket{\bar
\Phi^i(\lambda_L)}, \nonumber\\ \ket{\Psi_e (\lambda_R)} &=& \ket{
z_i}\otimes \ket{ \Phi^i(\lambda_R)},
\end{eqnarray}
describe excited eigenstates of the Hamiltonian $H$,
\begin{eqnarray}
H \ket{\bar \Psi_e (\lambda_L)} = \bar E_e(\lambda_L) \ket{\Psi_e
(\lambda_L)}, \nonumber\\
H \ket{\Psi_e (\lambda_R)} =  E_e(\lambda_R) \ket{\Psi_e
(\lambda_R)},
\end{eqnarray}
with eigenenergies $\bar E_e(\lambda_L)\geq E_0(\lambda_L)$ and
$E_e(\lambda_R) \geq \bar E_0(\lambda_R)$, which can be found from
the equations
\begin{eqnarray} \label{HLR2}
\bar H_{\Phi}^i(\lambda_L)\ket{\bar \Phi^i(\lambda_L)} = \bar
E_e(\lambda_L)
\ket{\bar \Phi^i(\lambda_L)}, \nonumber\\
  H_{\Phi}^i(\lambda_R)\ket{
\Phi^i(\lambda_R)} =  E_e(\lambda_R) \ket{\Phi^i(\lambda_R)}.
\end{eqnarray}

Since $H$ is a continuous function of $\lambda$ by definition,
$H_{\Phi}^i$ and $\bar H_{\Phi}^i$ should also be continuous
functions. Thus, as $\lambda_L \rightarrow \lambda_R$ ($\delta
\lambda \rightarrow 0$) we have: $H_{\Phi}^i(\lambda_L) \rightarrow
H_{\Phi}^i(\lambda_R)$ and $\bar H_{\Phi}^i(\lambda_L) \rightarrow
\bar H_{\Phi}^i(\lambda_R) $. This means that
$\ket{\Phi^i(\lambda_L)} \rightarrow \ket{\Phi^i(\lambda_R)}$ and
$\ket{\bar \Phi^i(\lambda_L)} \rightarrow \ket{\bar
\Phi^i(\lambda_R)}$. Comparing the first line of (\ref{HLR1}) and
the second line of (\ref{HLR2}), and vice versa. we find that at
$\lambda_L= \lambda_R = \lambda_0$, \be \label{ELR2} E_0(\lambda_0)
= E_e(\lambda_0), \qquad \bar E_0(\lambda_0) = \bar E_e(\lambda_0).
\ee Since $E_0(\lambda_0)$ and $\bar E_0(\lambda_0)$ both represent
the ground state energy at $\lambda_0$, they should be equal,
therefore,
\begin{eqnarray}
E_0(\lambda_0) = \bar E_e(\lambda_0) = \bar E_0(\lambda_0) =
E_e(\lambda_0).
\end{eqnarray}
We note that the energy levels $E_0$ and $\bar E_0$ correspond to
the eigenstates $\ket{\Psi_0}$ and $\ket{\bar \Psi_0}$ distinguished
by opposite directions of the $i$th spin  (see Eqs.~(\ref{PsiLR0})).
This means that there needs to be a degeneracy at
$\lambda=\lambda_0$, in contradiction with the
original assumptions.\\

\noindent (iii) $\ket{\psi_j(\lambda_R)} = \ket{z_j}$. The ground
states at $\lambda = \lambda_L$ and $\lambda = \lambda_R$ are
determined by
\begin{eqnarray} \label{PsiLR1}
\ket{\Psi_0 (\lambda_L)} &=& \ket{z_i}\otimes
\ket{\Phi^i(\lambda_L)}, \nonumber\\ \ket{\Psi_0 (\lambda_R)} &=&
\ket{ z_j}\otimes \ket{ \Phi^j(\lambda_R)}.
\end{eqnarray}
Since $H(\lambda)$ is continuous, one can use Taylor expansion
\begin{equation}
H(\lambda_L) = H(\lambda_R)  - \left[ \frac{\delta
H(\lambda)}{\delta \lambda}\right]_{\lambda = \lambda_R} \, \delta
\lambda.
\end{equation}
Applying the perturbation theory we find the ground state function
at $\lambda = \lambda_R$,
\begin{eqnarray} \label{PsiL1}
\ket{\Psi_0 (\lambda_L)} = \ket{\Psi_0 (\lambda_R)} - \delta \lambda
\, \ket{\Psi_0' (\lambda_R)},
\end{eqnarray}
where
\begin{eqnarray}
\ket{\Psi_0' (\lambda)} = \sum_{n > 0} \frac{\braket{\Psi_n(\lambda)
| \frac{\delta H(\lambda)}{\delta \lambda} | \Psi_0(\lambda)} } {
E_0(\lambda) - E_n(\lambda)}.
\end{eqnarray}
A similar perturbation procedure can be applied to find the function
$\ket{\Phi^i(\lambda_L)}$,
\begin{equation}
\ket{\Phi^i(\lambda_L)} = \ket{\Phi^i(\lambda_R)} - \delta \lambda
\ket{\Phi^{i\prime}(\lambda_R)}.
\end{equation}
Thus, for the function $\ket{\Psi_0(\lambda_L)}$ (see the first
Eq.~(\ref{PsiLR1})) we obtain
\begin{eqnarray} \label{PsiL2}
\ket{\Psi_0(\lambda_L)} = \ket{z_i}\otimes\ket{\Phi^i(\lambda_R)} -
\delta \lambda \ket{z_i}\otimes  \ket{\Phi^{i\prime}(\lambda_R)}.
\end{eqnarray}
Comparing with (\ref{PsiL1}), taking into account (\ref{PsiLR1}), to
the zeroth order perturbation, we obtain
\begin{equation}
\ket{z_i}\otimes\ket{\Phi^i(\lambda_R)} =  \ket{ z_j}\otimes \ket{
\Phi^j(\lambda_R)},
\end{equation}
which can be true if
\begin{eqnarray}
\ket{\Phi^i(\lambda_R)} &=& \ket{z^j}\otimes
\ket{\Phi^{ij}(\lambda_R)}, \nonumber\\
\ket{\Phi^j(\lambda_R)} &=& \ket{z^i}\otimes
\ket{\Phi^{ij}(\lambda_R)},
\end{eqnarray}
where $\ket{\Phi^{ij}(\lambda_R)}$ belongs to the Hilbert space
$S_{ij}$ of all qubits except qubits $i$ and $j$. Therefore,
\begin{equation}
\ket{\Psi_0(\lambda_R)} = \ket{z_i}\otimes \ket{z_j} \otimes
\ket{\Phi^{ij}(\lambda_R)}.
\end{equation}
Comparing with the first line in (\ref{PsiLR1}) it is clear that
$\braket{\sigma^z_i}$ does not change from $\lambda_L$ to
$\lambda_R$.

Looking back at all three cases, we see that for (i) and (iii),
$\braket{\sigma^z_i}$ does not change with $\lambda$ as the theorem
states, while (ii) contradicts with the conditions of the theorem,
namely the non-degeneracy of the ground state. $\square$

\section{Linear susceptibility}
To calculate susceptibility using the non-degenerate perturbation
theory (at small external biases $h_j$) we write the Hamiltonian as
$
  H = H_0 - \sum_j h_j \sigma^z_j,
$
where $H_0$ is the unperturbed Hamiltonian with eigenstates
$\ket{\Psi_n}$ and eigenvalues $E_n$.
To the first order perturbation in $h_j$, the ground state of $H$ is
given by
 \be \label{Psi0}
 \ket{\tilde{\Psi}_0} = \ket{\Psi_0} - \sum_j h_j \sum_{n>0} {\braket{\Psi_n|\sigma^z_j|\Psi_0} \over E_0-E_n}\ket{\Psi_n}.
 \ee
One therefore obtains
 \ba
 \braket{\tilde{\Psi}_0|\sigma^z_i|\tilde{\Psi}_0} &=& \braket{\Psi_0|\sigma^z_i|\Psi_0}
 \nonumber\\
 &-& \sum_j h_j \sum_{n>0} \frac{\braket{\Psi_0|\sigma^z_j|\Psi_n}\braket{\Psi_n|\sigma^z_i|\Psi_0}} {E_0-E_n}  \nonumber\\
 &-& \sum_j h_j \sum_{n>0} \frac{\braket{\Psi_0|\sigma^z_i|\Psi_n}\braket{\Psi_n|\sigma^z_j|\Psi_0}} {E_0-E_n}.  \nonumber
 \ea
The cross-susceptibility can be written as
 \ba
 \chi_{ij} &=& {\partial \braket{\sigma^z_i} \over \partial h_j}
 = \sum_{n>0} \frac{\braket{\Psi_0|\sigma^z_j|\Psi_n}\braket{\Psi_n|\sigma^z_i|\Psi_0}}
 {E_n-E_0}\nonumber\\
  &+& \sum_{n>0} \frac{ \braket{\Psi_0|\sigma^z_i|\Psi_n}\braket{\Psi_n|\sigma^z_j|\Psi_0}}
 {E_n-E_0}.
 \ea

\end{document}